# Linear-$T$ resistivity and change in Fermi surface at the pseudogap critical point of a high-$T_c$ superconductor

R. Daou [1*], Nicolas Doiron-Leyraud [1*], David LeBoeuf [1], S.Y. Li [1], Francis Laliberté [1],

Olivier Cyr-Choinière [1], Y.J. Jo [2], L. Balicas [2], J.-Q. Yan [3], J.-S. Zhou [3],

J.B. Goodenough [3] & Louis Taillefer [1,4†]

1 Département de physique and RQMP, Université de Sherbrooke, Sherbrooke, Québec
J1K 2R1, Canada

2 National High Magnetic Field Laboratory, Florida State University, Tallahassee,
Florida 32306, USA

3 Texas Materials Institute, University of Texas at Austin, Austin, Texas 78712, USA

4 Canadian Institute for Advanced Research, Toronto, Ontario M5G 1Z8, Canada

* These authors contributed equally to this work

† e-mail: louis.taillefer@physique.usherbrooke.ca

**A fundamental question of high-temperature superconductors is the nature of the pseudogap phase which lies between the Mott insulator at zero doping and the Fermi liquid at high doping $p$ (refs. 1, 2). Here we report on the behaviour of charge carriers near the zero-temperature onset of that phase, namely at the critical doping $p^*$ where the pseudogap temperature $T^*$ goes to zero, accessed by investigating a material in which superconductivity can be fully suppressed by a steady magnetic field. Just below $p^*$, the normal-state resistivity and Hall coefficient of $La_{1.6-x}Nd_{0.4}Sr_xCuO_4$ are found to rise simultaneously as the temperature drops below $T^*$, revealing a change in the Fermi surface with a large associated drop in**

**conductivity. At _p_\*, the resistivity shows a linear temperature dependence as _T_ → 0, a typical signature of a quantum critical point[3]. These findings impose new constraints on the mechanisms responsible for inelastic scattering and Fermi-surface transformation in theories of the pseudogap phase[1,4,5,6,7,8].**

At low hole doping _p_, high-$T_c$ superconductors are doped Mott insulators, strongly correlated metals characterized by a low carrier density _n_ equal to the concentration of doped holes[1]. Indeed, Hall effect measurements on La$_{2-x}$Sr$_x$CuO$_4$ (LSCO) at _x_ = _p_ < 0.05 yield a Hall number $n_H \equiv V / eR_H$ equal to _p_ at low temperature[9], where $R_H$ is the Hall coefficient, _e_ is the electron charge and _V_ is the volume per Cu atom. At high doping, however, these materials are Fermi liquids, metals characterized by a well-defined coherent three-dimensional Fermi surface[10] and a resistivity ρ that grows quadratically with temperature[11]: ρ ~ $T^2$. In that regime, the Fermi surface is a large cylinder containing 1 + _p_ holes[10], so the carrier density is high, given by _n_ = 1 + _p_. At _p_ ≈ 0.25, low-temperature measurements on Tl$_2$Ba$_2$CuO$_{6+y}$ (Tl-2201) yield $n_H$ = 1 + _p_ (ref. 12). These findings naturally beg the following question: How do the electrons in copper oxide superconductors go from one state to the other?

This is intimately tied to the question of what is the nature of the "pseudogap phase", this enigmatic region of the doping phase diagram present in all high-$T_c$ superconductors below a crossover temperature _T_\* (ref. 2). Here we investigate the _T_ = 0 onset of this pseudogap phase by measuring the transport properties of La$_{1.6-x}$Nd$_{0.4}$Sr$_x$CuO$_4$ (Nd-LSCO), a material whose relatively low maximal $T_c$ makes it possible to suppress superconductivity entirely with a steady magnetic field.

In Fig. 1, we show the normal-state resistivity ρ(_T_) of Nd-LSCO at a doping _p_ = 0.20. Above a temperature _T_\* = 80 K, ρ(_T_) exhibits the linear temperature dependence characteristic of all hole-doped copper oxides. Below that temperature, it deviates upwards and develops an upturn visible even in zero field (see Supplementary

Fig. S1), with a minimum at $T_{min}$ = 37 K > $T_c$ = 20 K, in excellent agreement with early data in zero field[13]. By applying a magnetic field of 35 T, we were able to track the upturn in $\rho(T)$ down to 1 K, thus revealing a pronounced rise at low temperature (Fig. 1).

The absence of magneto-resistance (see Supplementary Fig. S1) implies that the magnetic field simply serves to remove superconductivity and reveal the unaltered behaviour of the underlying normal state down to $T \approx 0$. The evolution with temperature is perfectly smooth, indicating a crossover as opposed to a transition. Most significantly, $\rho(T)$ saturates at low temperature (see Fig. 2a). This shows that the ground state is a metal and not an insulator, and that $T^*$ therefore marks the onset of a crossover from one metallic state to another. Note that the loss of conductivity is substantial, by a factor of approximately $\rho_0 / \rho(T{\rightarrow}0)$ = 5.8, where $\rho_0$ = 245 $\mu\Omega$ cm is the resistivity measured at $T$ = 1 K and $\rho(T{\rightarrow}0)$ = 42 $\mu\Omega$ cm is the value extrapolated linearly to $T$ = 0 from above $T^*$.

We identify $T^*$ as the onset of the pseudogap phase, following the standard definition: the temperature below which the in-plane resistivity $\rho_{ab}(T)$ starts to deviate from its linear-$T$ behaviour at high temperature[2,14]. (Note that the deviation can be either upwards, as in LSCO, or downwards, as in YBa$_2$Cu$_3$O$_y$ (YBCO) (ref. 14), depending on the relative magnitude of inelastic and elastic (disorder) scattering at $T^*$. In YBCO, the copper oxide material with the lowest disorder scattering, the loss of inelastic scattering below $T^*$ is a much larger relative effect than in LSCO, hence the drop in $\rho_{ab}(T)$.) In Fig. 3, we plot $T^*$ as a function of doping in a $p - T$ phase diagram. Note that the magnitude of $T^*$ in Nd-LSCO is comparable to that found in other hole-doped copper oxides, pointing to a common origin (see Supplementary Fig. S2 for a comparison with LSCO).

In Fig. 4, we present the Hall coefficient $R_H(T)$ measured on the same crystal (with $p$ = 0.20), and compare it directly to $\rho(T)$. Both coefficients are seen to rise simultaneously, with $T_{min}$ the coincident onset of their respective upturns. This is strong evidence that the cause of both upturns is a modification of the Fermi surface.

Let us now look at a slightly higher doping. Figs. 1 and 4 respectively show $\rho(T)$ and $R_H(T)$ measured on a second crystal, with $p = 0.24$. The low-temperature behaviour has changed: $\rho(T)$ shows no sign of an upturn and $R_H(T)$ is now constant below 25 K, extrapolating to $R_H = + 0.45 \pm 0.05$ mm$^3$ / C as $T \to 0$. The corresponding Hall number is $n_H = 1.3 \pm 0.15$, in good agreement with the carrier density $n = 1 + p = 1.24$ expected for a large Fermi cylinder, and quantitatively consistent with measurements on Tl-2201 at $p = 0.26$ and $T \to 0$, where $n_H = 1.3$ (ref. 12). By comparison, at $p = 0.20$, the magnitude of $R_H$ at $T \to 0$ yields $n_H = 0.3 \pm 0.05$. The change in the Hall number at $T \to 0$ between $p = 0.24$ and $p = 0.20$ is therefore $\Delta n_H = 1.0 \pm 0.2$ hole per Cu atom. If the Hall number is interpreted as a carrier density, these values are consistent with a crossover from a metal with a large hole-like Fermi surface at $p^*$ (where $n = 1 + p$) to a metal with a low density of holes below $p^*$ (where $n \approx p$).

In contrast to $p = 0.20$, the electrical resistivity at $p = 0.24$ shown in Fig. 2b displays a monotonic temperature dependence down to 1 K, linear as $T \to 0$. The absence of any anomaly demonstrates that $T^* = 0$ at that doping. Therefore the critical doping $p^*$ where the pseudogap line ends is located between $p = 0.20$ and $0.24$, inside the region where superconductivity exists in zero field. For definiteness, in Fig. 3 we set it at $p^* = 0.24$, although it could be slightly lower.

As shown in Fig. 2b, not only is the in-plane resistivity $\rho_{ab}(T)$ linear as $T \to 0$ at $p = 0.24$, but so is the out-of-plane resistivity $\rho_c(T)$. Moreover, the fact that $R_H(T)$ is flat at low temperature implies that the cotangent of the Hall angle, cot $\theta_H(T) \sim \rho_{ab}(T) / R_H(T)$, is also linear at low temperature. We infer that a single anomalous scattering process dominates the electron-electron correlations at low temperature at $p^*$ (or just above). This shows that the Fermi-liquid behaviour observed at $p = 0.3$ (in LSCO), where $\rho_{ab}(T) \sim T^2$ below $T \approx 50$ K (ref. 11), breaks down just before the onset of the pseudogap phase at $p^*$. This kind of "non-Fermi-liquid" behaviour, whereby $\rho(T) \sim T$ as $T \rightarrow 0$, has typically been observed in heavy-fermion metals at the quantum critical point where the onset temperature for antiferromagnetic order goes to zero[3]. It is also consistent with the marginal Fermi liquid description of cuprates[15].

In summary, our experimental findings offer compelling evidence that the pseudogap phase ends at a $T = 0$ critical point $p^*$ located below the onset of superconductivity (at $p_c \approx 0.27$), in agreement with previous but more indirect evidence from other hole-doped copper oxides[16]. Moreover, they impose two strong new constraints on theories of the pseudogap phase: 1) its onset below $p^*$ modifies the large Fermi surface characteristic of the overdoped metallic state; 2) quasiparticle scattering at $p^*$ is linear in temperature as $T \rightarrow 0$.

The existence of a quantum critical point is consistent with two kinds of theories of the pseudogap phase. The first kind invokes the onset of an order, with some associated broken symmetry (refs. 6, 7, 8). Because $T^*$ marks a crossover and not a sharp transition, this order is presumably short-range or fluctuating. In the electron-doped copper oxides, for example, the pseudogap phase has been interpreted as a fluctuating precursor of the long-range antiferromagnetic order which sets in at lower temperature[17], and the signatures of the pseudogap critical point in transport are similar to those found here: a linear-$T$ resistivity as $T \rightarrow 0$ (ref. 18) and a sharp change in $R_H(T = 0)$ (ref. 19). For Nd-LSCO and LSCO, an analogous scenario would be "stripe" fluctuations, as a precursor to the static spin and charge modulations observed at lower temperature[20]. Note that in Nd-LSCO at $p = 0.20$, the onset of the upturn in $\rho(T)$ and $R_H(T)$ at $T_{min} = 37$ K coincides with the loss of NQR intensity at $T_{NQR} = 40 \pm 6$ K (ref. 21) (see Fig. 4). In Nd-LSCO at $p = 0.15$, this so-called "wipeout" anomaly in NQR at $T_{NQR} = 60 \pm 6$ K (ref. 21) was shown to coincide with the onset of charge order measured via hard X-ray diffraction, at $T_{ch} = 62 \pm 5$ K (ref. 22) (see Fig. 3). Direct evidence of a charge modulation via resonant soft X-ray diffraction was reported recently for the closely-related material Eu-LSCO, with $T_{ch} = 70 \pm 10$ K at $p = 0.15$ (ref. 23), while $T_{NQR} = 60 \pm 6$ K (ref. 21) in Eu-LSCO at $p = 0.16$ (see Fig. 3). Clearly, the upturn in $\rho(T)$ is correlated with the onset of charge order in these two materials. While the correlation between $T_{NQR}$ and $T_{min}$ has been noted previously[20], the mechanism causing the upturn in $\rho(T)$ remained unclear. Our data shows that the mechanism is a change in Fermi surface, and the positive rise in $R_H(T)$ imposes a strong constraint on the topology of the resulting Fermi surface. An additional constraint comes from the fact that $R_H(T)$ drops to negative values near $p = 1/8$, not only in Nd-LSCO (ref. 13) and other materials with "stripe" order[24,25], but also in YBCO (ref. 26).

Recent calculations of the Fermi-surface reconstruction caused by stripe order are consistent with a negative $R_H$ near $p = 1/8$ in that spin stripes tend to generate an electron pocket in the Fermi surface[27]. Interestingly, charge stripes do not[27], and this might explain the positive rise in $R_H$ seen at higher doping, provided that stripe order involves predominantly charge order at high doping (in line with the fact that charge order sets in at a higher temperature than spin order[20,21]).

In the other kind of theories of the pseudogap phase, the critical point reflects a $T = 0$ transition from small hole pockets, characteristic of a doped Mott insulator, to a large hole pocket, without symmetry breaking[4,5]. Recent work suggests that the quasiparticle scattering rate above such a critical point may indeed grow linearly with temperature[28]. Although calculations are needed to confirm this, a change in carrier density from $n \approx p$ to $n = 1 + p$ would seem natural in this kind of scenario. However, it is more difficult to see what could cause the negative values of $R_H(T \rightarrow 0)$ near $p = 1/8$. It seems that stripe order or fluctuations would have to be invoked as a secondary instability inside the pseudogap phase, with an onset in doping that would be essentially simultaneous with $p*$ in the case of Nd-LSCO.

We end by comparing our results qualitatively with those of previous high-field studies on LSCO. The resistivity shows very similar features at high temperature: linear-$T$ above $T*$ (ref. 14) and an upturn below $T*$ (ref. 29). But the Hall coefficient of LSCO[30], on the other hand, has a more subtle and complex evolution with doping than that presented here for Nd-LSCO, which makes it harder to pinpoint $p*$ using the same criteria as we have used above. Nonetheless, it seems likely that the same fundamental mechanisms are responsible for both the linear-$T$ resistivity and the resistivity upturns, and for the onset of the pseudogap at $T*$, in both LSCO and Nd-LSCO.

**METHODS**

Single crystals of $La_{2-y-x}Nd_ySr_xCuO_4$ (Nd-LSCO) were grown with a Nd content $y = 0.4$ using a travelling float zone technique and cut from boules with nominal Sr concentrations $x = 0.20$ and $x = 0.25$. The actual doping $p$ of each crystal was estimated from its $T_c$ and $\rho(250$ K$)$ values compared with published data, giving $p = 0.20 \pm 0.005$ and $0.24 \pm 0.005$, respectively. The resistivity $\rho$ and Hall coefficient $R_H$ were measured at the NHMFL in Tallahassee in steady magnetic fields up to 35 T and in Sherbrooke in steady fields up to 15 T. The field was always applied along the $c$-axis. Neither $\rho$ nor $R_H$ showed any field dependence up to the highest fields. More details are available in the Supplementary Information.

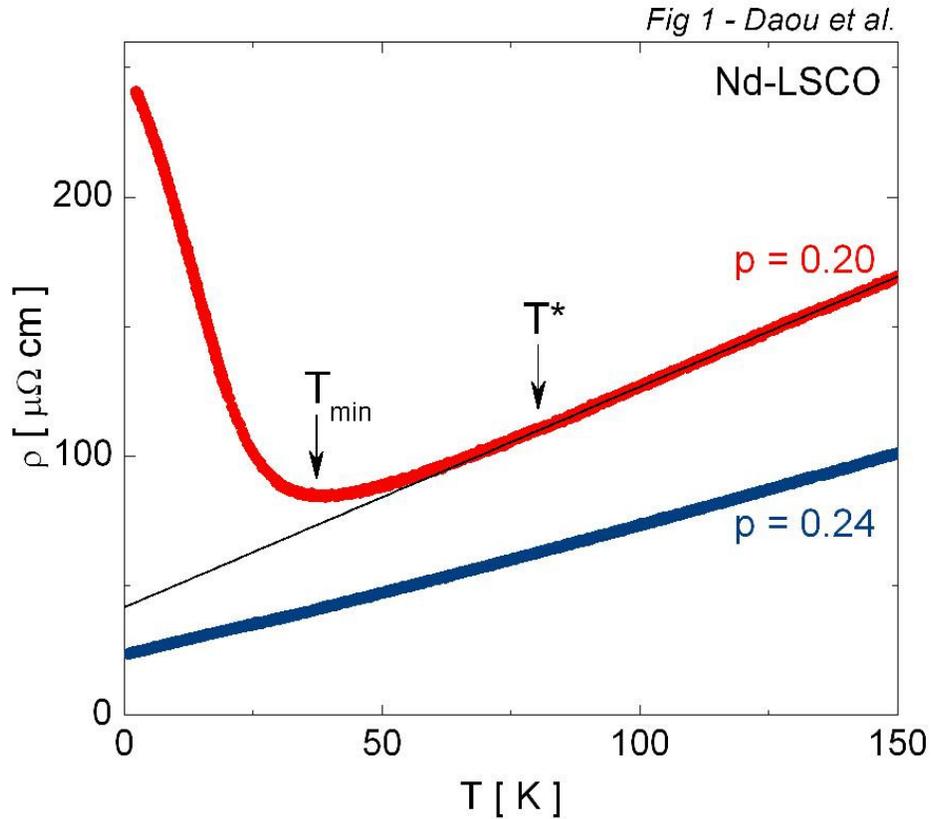

*Fig 1 - Daou et al.*

**Figure 1 | Normal-state resistivity.**

In-plane electrical resistivity $\rho(T)$ of Nd-LSCO as a function of temperature, at $p = 0.20$ and $p = 0.24$, measured in a magnetic field strong enough to fully suppress superconductivity [see Supplementary Information]. The black line is a linear fit to the $p = 0.20$ data between 80 and 300 K. Below a temperature $T^* = 80$ K, $\rho(T)$ deviates from its linear-$T$ behaviour at high temperature and develops a pronounced upturn at low temperature, with a minimum at $T_{min} = 37$ K. By contrast, $\rho(T)$ at $p = 0.24$ shows no upturn down to the lowest temperature.

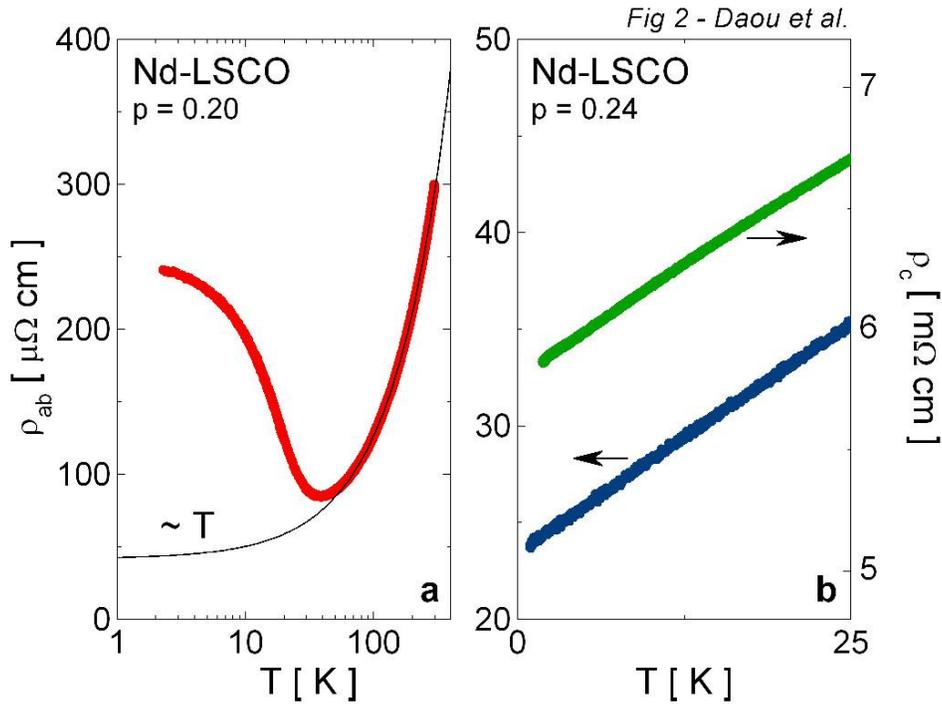

**Figure 2 | In-plane and out-of-plane resistivities at low temperature.**

**a)** Semi-log plot of the in-plane resistivity of Nd-LSCO at $p = 0.20$. The black line is a linear fit above 80 K. This shows that, after a rapid rise, $\rho_{ab}(T)$ saturates at low temperature, in contrast with the weak logarithmic divergence observed in LSCO at $p < 0.16$ (ref. 29). **b)** Temperature dependence of the normal-state electrical resistivity of Nd-LSCO at $p = 0.24 \approx p^*$, in the low-temperature regime. Both the in-plane resistivity $\rho_{ab}$ and the out-of-plane resistivity $\rho_c$ show a linear temperature dependence down to the lowest measured temperature.

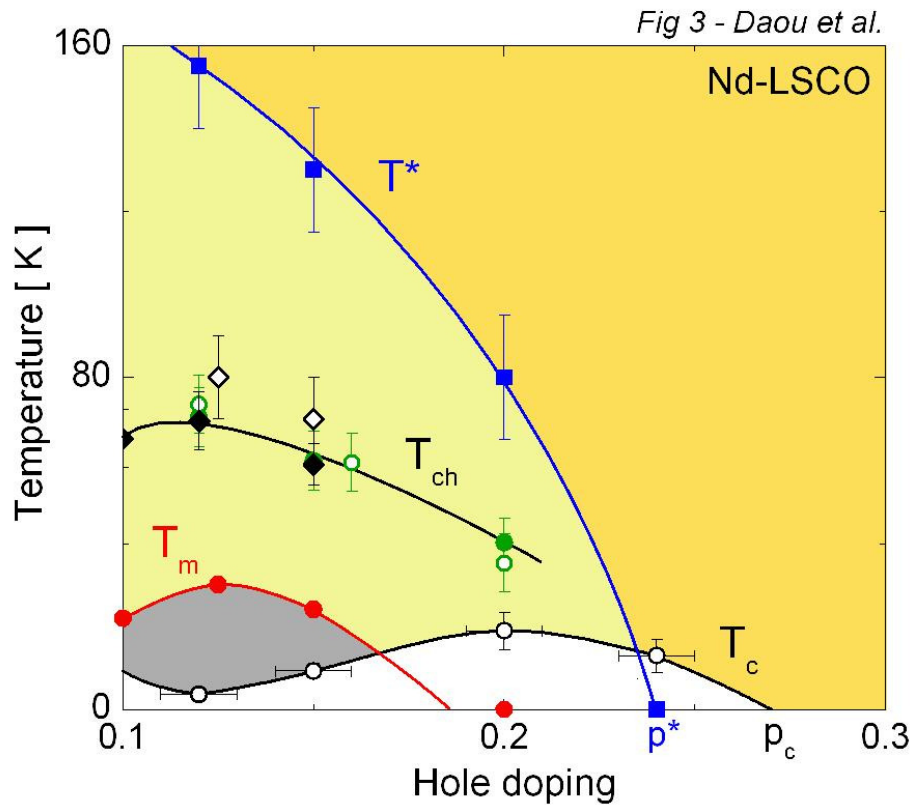

**Figure 3 | Phase diagram.**

Temperature-doping phase diagram of Nd-LSCO showing the superconducting phase below $T_c$ [open black circles] and the pseudogap region delineated by the crossover temperature $T^*$ [blue squares]. Also shown is the region where static magnetism is observed below $T_m$ [red circles] and charge order is detected below $T_{ch}$ [black diamonds and green circles]. These onset temperatures are respectively defined as the temperature below which: 1) the resistance is zero; 2) the in-plane resistivity $\rho_{ab}(T)$ deviates from its linear dependence at high temperature; 3) an internal magnetic field is detected by zero-field muon spin relaxation (μSR); 4) charge order is detected by either X-ray diffraction or NQR.

All lines are a guide to the eye. Values of $T_c$ and error bars are given in Supplementary Information. $T^*$ is obtained from a fit to the $\rho_{ab}(T)$ data of ref. 20 for $p = 0.12$ and $p = 0.15$, and that reported here for $p = 0.20$ and 0.24 (See Fig. 1 and S2, and Supplementary Information for fits and error bars). The blue line above $p = 0.20$ is made to end at $p = 0.24$, thereby defining the critical doping where $T^*$ goes to zero as $p^* = 0.24$. Experimentally, this point must lie in the range $0.20 < p^* \leq 0.24$, since $\rho(T)$ remains linear down to the lowest temperature at $p = 0.24$ (see Fig. 2b). $T_m$ is obtained from the µSR measurements of ref. 31. The red line is made to end below $p = 0.20$, as no static magnetism was detected at $p = 0.20$ down to $T = 2$ K. $T_{ch}$ is obtained from hard X-ray diffraction on Nd-LSCO [full black diamonds and error bars; ref. 22] and from resonant soft X-ray diffraction on Eu-LSCO [open diamonds and error bars; ref. 23]. The onset of charge order has been found to coincide with the wipeout anomaly in NQR at $T_{NQR}$, reproduced here from ref. 21 (error bars quoted therein) for Nd-LSCO [closed green circles] and Eu-LSCO [open green circles].

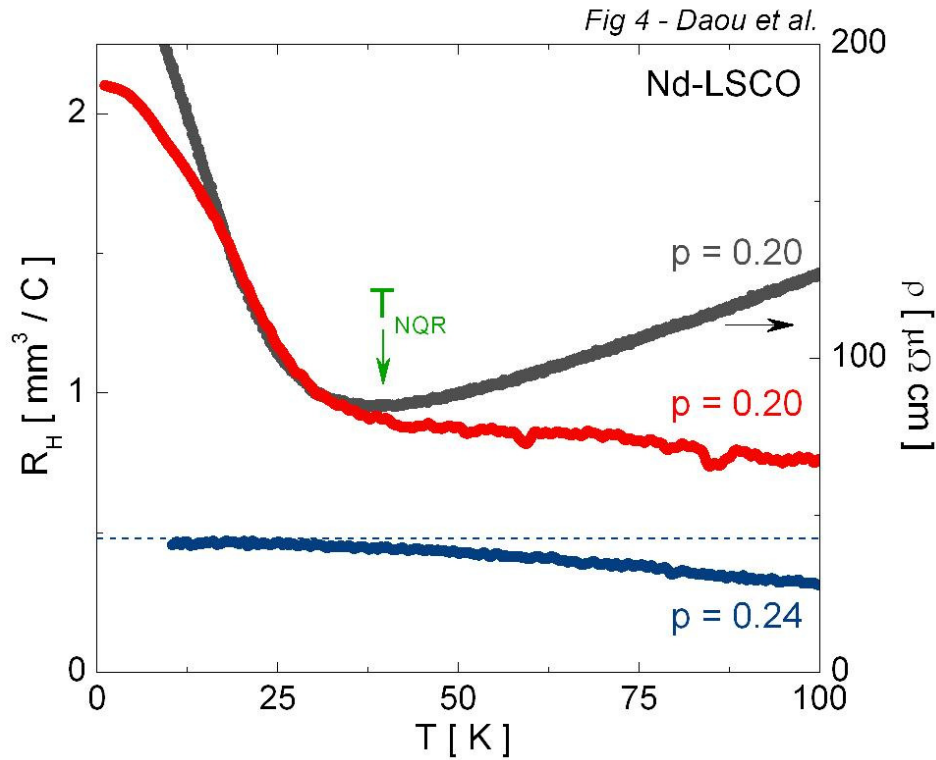

**Figure 4 | Normal-state Hall coefficient.**

Hall coefficient $R_H(T)$ of Nd-LSCO as a function of temperature for $p = 0.20$ and 0.24, measured in a magnetic field of 15 T. Below 12 K, the 0.20 data are in 33 T, a magnetic field strong enough to fully suppress superconductivity [see Supplementary Information]. The dashed blue horizontal line is the value of $R_H$ calculated for a large cylindrical Fermi surface enclosing $1 + p$ holes, namely $R_H = V / e\,(1 + p)$, at $p = 0.24$. At $p = 0.20$, the rise in $R_H(T)$ at low temperature signals a modification of that large Fermi surface. The upturn is seen to coincide with a simultaneous upturn in $\rho(T)$ (reproduced in grey from Fig. 1) and with the onset of charge order at $T_{NQR}$ as detected by NQR (see text and ref. 21).

Supplementary information accompanies this paper on www.nature.com/naturephysics.

**Acknowledgements** We thank K. Behnia, A. Chubukov, P. Coleman, Y.B. Kim, S.A. Kivelson, G. Kotliar, K. Haule, G.G. Lonzarich, A.J. Millis, M.R. Norman, C. Proust, T.M. Rice, S. Sachdev, T. Senthil, H. Takagi and A.-M.S. Tremblay for helpful discussions, and J. Corbin for his assistance with the experiments. LT acknowledges support from the Canadian Institute for Advanced Research and funding from NSERC, FQRNT, and a Canada Research Chair. LB was supported by NHMFL-UCGP and YJJ by the NHMFL-Schuller fellow program. JSZ and JBG were supported by an NSF grant. The NHMFL is supported by an NSF grant and the State of Florida.

**Author Information** Reprints and permission information is available online at http://npg.nature.com/reprintsandpermissions. Correspondence and requests for materials should be addressed to L.T.

---

# Supplementary Information for

# "Linear-*T* resistivity and change in Fermi surface at the pseudogap critical point of a high-*T*c superconductor"

R. Daou [1], Nicolas Doiron-Leyraud [1], David LeBoeuf [1], S.Y. Li [1], Francis Laliberté [1],

Olivier Cyr-Choinière [1], Y.J. Jo [2], L. Balicas [2], J.-Q. Yan [3], J.-S. Zhou [3],

J.B. Goodenough [3] & Louis Taillefer [1,4]

*1 Département de physique and RQMP, Université de Sherbrooke, Sherbrooke, Québec J1K 2R1, Canada*

*2 National High Magnetic Field Laboratory, Florida State University, Tallahassee, Florida 32306, USA*

*3 Texas Materials Institute, University of Texas at Austin, Austin, Texas 78712, USA*

*4 Canadian Institute for Advanced Research, Toronto, Ontario M5G 1Z8, Canada*

## METHODS

**In-plane samples**.  The two samples of Nd-LSCO used for in-plane transport were grown using a traveling float zone technique in an image furnace, as described in ref. 1. The nominal Sr concentration for the two growths was $x = 0.20$ and $x = 0.25$, respectively. The actual doping $p$ of small crystals cut out of the resulting large boules is expected to be $p = x \pm 0.01$. The physical dimensions of the samples cut out of the single-crystal boules were measured using an optical microscope and are shown in Table 1.

**Table 1**

| Sample (in-plane) | Length, $L$ [mm] | Width, $w$ [mm] | Thickness, $t$ [mm] |
|---|---|---|---|
| Nd-LSCO x=0.20 | 1.51 ± 0.05 | 0.50 ± 0.02 | 0.64 ± 0.02 |
| Nd-LSCO x=0.25 | 2.50 ± 0.05 | 0.51 ± 0.02 | 0.51 ± 0.02 |

**c-axis sample**.  One sample of Nd-LSCO was used for $c$-axis resistivity, shown in Fig. 2b. It was cut from the same $x = 0.25$ boule as the $x = 0.25$ in-plane sample. It had a resistive $T_c = 17 \pm 0.5$ K in zero field.

**Superconducting transition temperature $T_c$**. The superconducting transition temperature $T_c$ of our Nd-LSCO in-plane samples was determined via magnetic susceptibility measurements in a SQUID magnetometer. $T_c$ is defined as the midpoint of the transition and the width is that of the susceptibility drop between 90 % and 10 %. In Table 2, these $T_c$ values are compared with $T_c$ defined as the temperature where the resistivity goes to zero.

**Table 2**

| Sample (in-plane) | $T_c$ [K] (midpoint) | $T_c$ [K] (width) | $T_c$ [K] ($\rho = 0$) |
|---|---|---|---|
| Nd-LSCO x=0.20 | 19 | 9 | 20.5 ± 0.5 |
| Nd-LSCO x=0.25 | 13 | 8 | 17.0 ± 0.5 |

**Hole doping $p$**. By comparing the $T_c$ values and the absolute values of the resistivity (at 250 K) in each of our two in-plane samples (in bold) with previous measurements on Nd-LSCO and LSCO (for which we assume $p = x$), we arrive at an estimate of the doping, given in Table 3.

**Table 3**

| Sample | x | Tc ($\rho = 0$) | $\rho$ (T = 250 K) [$\mu\Omega$ cm] | p | Ref. |
|---|---|---|---|---|---|
| Nd-LSCO | 0.20 | 20 ± 0.5 | 232 ± 20 | 0.20 | [2,3] |
| LSCO | 0.20 | N/A | 210 ± 20 | 0.20 | [3] |
| LSCO | 0.20 | N/A | 230 ± 20 | 0.20 | [4] |
| **Nd-LSCO** | **0.20** | **20.5 ± 0.5** | **237 ± 20** | **0.20 ± 0.005** | **This work** |
| LSCO | 0.22 | N/A | 192 ± 20 | 0.22 | [4] |
| **Nd-LSCO** | **0.25** | **17.0 ± 0.5** | **164 ± 15** | **0.24 ± 0.005** | **This work** |
| Nd-LSCO | 0.25 | 7 ± 0.5 | 132 ± 13 | 0.25 | [2,3] |
| LSCO | 0.25 | N/A | 116 ± 12 | 0.25 | [3] |

**Contacts**. Electrical contacts on the Nd-LSCO samples were made to the crystal surface using Epo-Tek H20E silver epoxy. This epoxy was cured for 5 min at 180 C, then annealed at 500 C in flowing oxygen for 1 hr so that the silver diffused into the surface. This resulted in contact resistances of less than 0.1 $\Omega$ at room temperature. The contacts were wrapped around all four sides of the sample. In addition, the current contacts covered the end faces. Hall contacts were placed opposite each other in the middle of the samples, extending along the length of the *c*-axis, on the sides. The uncertainty in the quoted length (between contacts) of the sample, and hence the geometric factor, reflects the width of the voltage contacts.

**Magnetic field direction**. In all measurements, the magnetic field was applied along the *c*-axis of the sample.

**Measurements of resistivity and Hall coefficient**. The resistivity $\rho(T) \equiv R_{xx}\, w\, t\, /\, L$ and Hall coefficient $R_H(T) \equiv R_{xy}\, t\, /\, H$ of each in-plane sample were measured using the standard six-terminal AC technique. A resistance bridge or a lock-in amplifier was used to measure the resistance. Field reversal was used to obtain the symmetric and anti-symmetric parts of the voltages, accounting for any misalignment of the contacts. Therefore, the longitudinal ($R_{xx}$) and transverse ($R_{xy}$) resistances were obtained as follows:

$$R_{xx} = (\,R(B) + R(\text{-}B)\,)\,/\,2 \quad \text{and} \quad R_{xy} = (R(B) - R(\text{-}B)\,)\,/\,2.$$

Low-field measurements were performed in Sherbrooke, in the temperature range 4-300 K using a steady magnetic field of up to 15 T. High-field measurements on the same samples were performed at the NHMFL in Tallahassee, in a helium-3 refrigerator in DC fields of up to 33 T ($R_{xy}$ for the $x = 0.20$ sample and $R_{xx}$ for the $x = 0.25$ sample) and 35 T ($R_{xx}$ for the $x = 0.20$ sample).

The resistivity $\rho_c(T)$ of the $c$-axis sample shown in Fig. 2b was measured at the NHMFL using a standard four-terminal technique, in a field of 30 T.

No dependence on magnetic field was observed in either the resistance $R_{xx}$ or the Hall coefficient $R_H = t\, R_{xy}\, /\, B$. Fig. S1 shows the separate data taken at $H = 0$ and $H = 15$ T in Sherbrooke and at $H = 35$ T in Tallahassee on sample $x = 0.20$ ($p = 0.20$); one can see that there is negligible magneto-resistance (in the normal state).

Hall effect data taken at the NHMFL (up to 33 T) on the $x = 0.20$ sample had a low signal-to-noise ratio. The data shown in Fig. 4 was smoothed by a running average of width 100 data points (taken very densely). In addition, a small offset of 10 % was recorded between the Sherbrooke data at 15 T and the NHMFL data at 33 T in the temperature region of overlap (15-25 K). In Fig. 4, the NHMFL data was normalized to match the lower-noise Sherbrooke results. This is justified given the absence of field dependence in both $R_{xx}$ and $R_H$ up to 35 T.

**Determination of $T^*$.** The pseudogap temperature $T^*$ is defined as the point where the resistivity deviates from its high-temperature linear behavior. In Fig. S2, we show how it can be obtained readily by plotting $\rho(T) - (\rho_0 + AT)$ versus temperature, where ($\rho_0 + AT$) is a fit to the high-temperature region. The $T^*$ values thus obtained are plotted vs doping in Fig. 3. Data on LSCO at $p = 0.14$ (from ref. 4) is also shown for comparison.

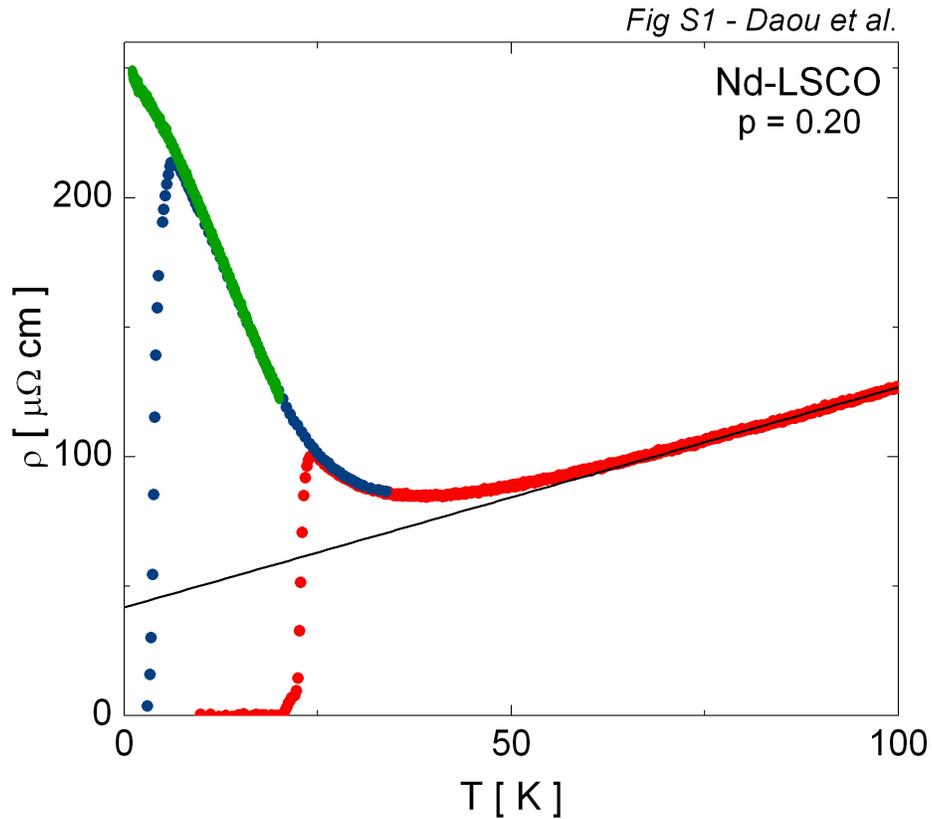

**Figure S1 | Electrical resistivity of Nd-LSCO at *p* = 0.20.**

Temperature dependence of the in-plane electrical resistivity, $\rho_{ab}(T)$, for the in-plane Nd-LSCO sample with $x = 0.20$ ($p = 0.20$), at three different magnetic fields: $H = 0$ (red), $H = 15$ T (blue), and $H = 35$ T (green). The fact that all three curves overlap almost perfectly (in the normal state) shows that there is negligible magneto-resistance. This implies that the effect of a magnetic field is simply to remove superconductivity, without altering the underlying normal state. The data shown in Fig. 1 and Fig. 2a includes all three sets of data.

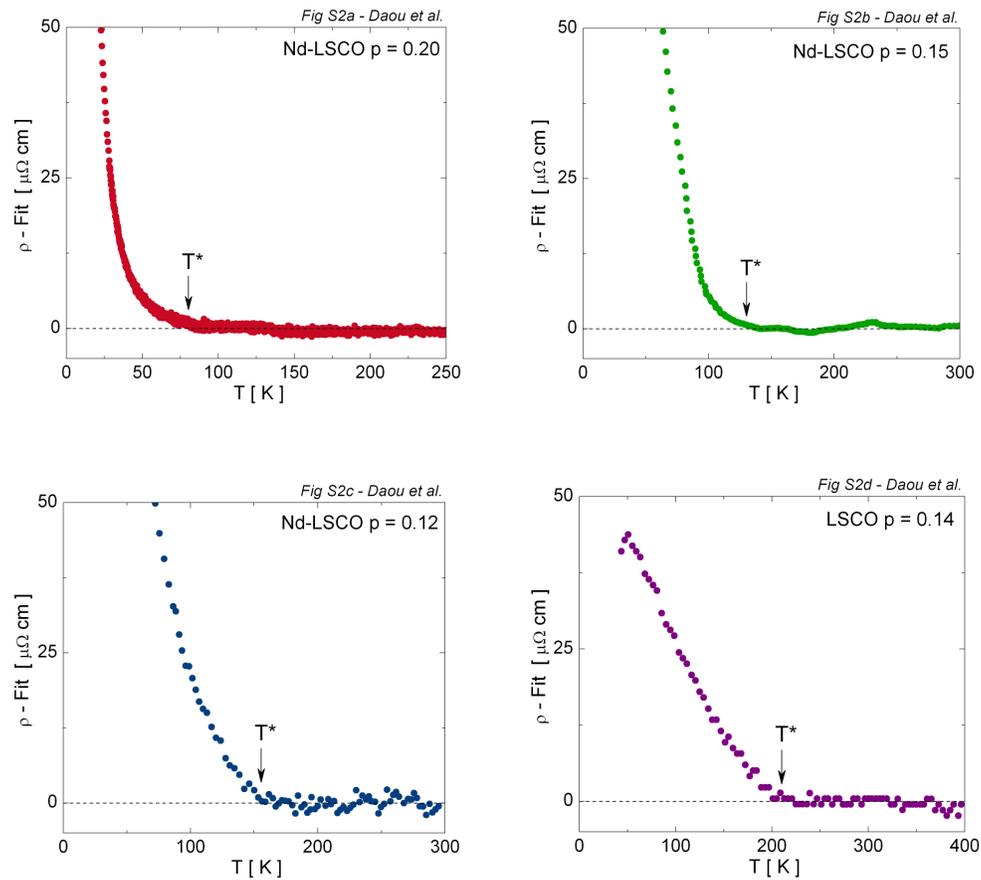

**Figure S2 | Pseudogap temperature in Nd-LSCO and LSCO.**

Resistivity $\rho(T)$ minus a linear fit to the high-temperature data of the form $\rho_0 + AT$. **a)** Nd-LSCO at $p = 0.20$ (this work); **b)** Nd-LSCO at $p = 0.15$ (from ref. 2); **c)** Nd-LSCO at $p = 0.12$ (from ref. 2). This allows us to define the pseudogap temperature $T^*$ as the end of the linear-$T$ regime at high temperature. The values obtained are $T^* = 80$, 130, and 155 K, respectively, with an error bar of about ± 15 K in all cases. They are plotted on a $p - T$ phase diagram in Fig. 3. **d)** LSCO at $p = 0.14$ (from ref. 4). This shows that the onset of the pseudogap has a similar signature in the resistivity of LSCO and Nd-LSCO, with comparable $T^*$ values.

———————————————